\documentclass[10pt,conference]{IEEEtran}
\usepackage{times}
\usepackage{psfrag}
\usepackage{graphicx}
\usepackage{amsfonts}

\begin{document}
\title{Debugging Memory Issues In Embedded Linux: \\ 
	A Case Study}
\author{Partha Pratim Ray $\quad$ $\quad$ $\quad$ $\quad$ $\quad$ Ansuman Banerjee \\
	Haldia Institute of Technology $\quad$ $\quad$ Indian Statistical Institute}
\maketitle
\bibliographystyle{latex8}
\thispagestyle{empty}


\begin{abstract}
\noindent
Debugging denotes the process of detecting root causes of unexpected 
observable behaviors in programs, such as a program crash, an unexpected output value being produced or an assertion violation. Debugging of program errors is
a difficult task and often takes a significant amount of time in the software
development life cycle. In the context of embedded software, the probability of bugs is quite high. Due to requirements 
of low code size and less resource consumption, embedded softwares typically
do away with a lot of sanity checks during development time. This leads to high chance of errors being uncovered in the production code at run time. In this 
paper we propose a methodology for debugging errors in BusyBox, a de-facto 
standard for Linux in embedded systems. Our methodology works on top of 
Valgrind, a popular memory error detector and Daikon, an invariant analyzer. 
We have experimented with two published errors in BusyBox and report our findings 
in this paper.
\end{abstract}

\section{Introduction} \label{sec1}
\noindent
Embedded software and systems have come to dominate the way we 
interact with computer and computation in our everyday life. 
Validation and debugging of embedded systems is therefore invariably, 
and inextricably an issue of paramount importance in today's fast 
paced development process and stringent time to market deadlines. 
As a result the debugging problem for embedded software systems has 
aroused significant research interest in the academic and industry 
community.

\noindent
Today's programming languages and compiler technologies have 
reached a fairly high level of sophistication, thereby allowing 
the programmers to design large complex pieces of code in less 
time. Embedded software systems are designed with some additional objectives (e.g. low code size, low memory foot print etc.) as compared to normal software. In keeping with these objectives, embedded system softwares are often designed without sufficient sanity checking aids (e.g. exceptions, signal handlers, assertions etc.). As a result, the probability of bugs being uncovered at production time in embedded software is quite high.

\noindent
A software bug[4] is an error, flaw, mistake, undocumented feature that prevents it from behaving as intended (e.g. producing an incorrect result). Most bugs arise due to mistakes and errors made by the programmer, either in the program source code or its design, and a few are caused by compilers producing incorrect code. It is worthwhile to note that manifestation of a bug may be very different from the bug itself, thus the main task in debugging is to trace back the bug source from the manifestation of it. A good bug report should be able to take in a manifestation of a bug and locate the root cause.

\noindent
In this paper, we propose a methodology for debugging memory usage errors in embedded software. Our method involves performing memory checking and invariant
analysis on the program under test. Given the observable errors, we perform memory checking using Valgrind[3], and deduce a set of causes of the observed memory error. This is followed by an invariant analysis using Daikon[2], one of the oldest and most widely used invariant detectors. Given a buggy program and a test input that demonstrates the bug, we symbolically extract the set of invariants from the buggy program and a stable program (not having the bug) for the same specification and compare the invariants thus found. The result of this comparison is analyzed with respect to the bug manifestation and points to a set of possible causes of the bug.

\noindent
We employ our method on BusyBox[1], the de-facto standard for Linux in embedded devices. 
It provides many of the standard utilities but has a smaller code size. Researchers have analyzed 
BusyBox and reported 21 bugs in BusyBox [8].
Our objective in this paper is to locate the root causes of some of these 
bugs using memory checking and invariant analysis. In particular, we work 
with two of the 21 bugs, namely, {\em arp} and {\em top}, both of which 
result in {\em segmentation violation} on execution in BusyBox version 1.4.2. 

\noindent
Our methodology for root cause analysis works in two steps.
Firstly, we perform memory checking using Valgrind on the buggy version of BusyBox. 
Valgrind directly reports memory errors, pointing to the lines of source where 
memory access is faulty. Secondly we choose two versions of 
BusyBox, namely, the buggy one and another one which does not demonstrate the 
bug and perform invariant extraction using Daikon. A comparison of the invariants 
thus produced lead us to the behavioral anomaly between the two program versions.

\noindent
This paper is organized as follows: Section II presents related work. Section III presents an overview of BusyBox. Section IV presents 
experiments done using Valgrind, while Section V describes our experimentation done with Daikon. 

\section{Related work}\label{sec2}
\noindent
In a recent publication, Jeff H. Perkins[6] et. al has proposed a system that 
automatically patches errors in deployed softwares. In this approach, 
{\em Clearview}[6] is used to correct unknown errors in the commercial off-the-shelf
(COTS) softwares. As per its architecture, learning, monitoring, correlated 
invariant identification, candidate repair generation and candidate 
repair evolution phases perform invariant analysis and memory checking 
rigorously. Daikon[3] is used in its learning component to analyze the various 
invariants present in the code. HeapGuard and Determina Memory 
Firewall[5] help in monitoring phase incorporating the monitor to detect a 
failure and the failure location.

\noindent
The work [7] illustrates invariant analysis
by Daikon while performing mutation test. Mutation 
test basically measures the adequacy of a test suite 
by introducing artificial defects (mutations) into a test program. The 
assessment of the mutations are done by dynamic invariants. Different versions
 (with invariant checkers) originating from a given source program, are 
checked by JAVALANCE (relies on invariant detection engine of DAIKON) to
find the best survived mutation.

\noindent
One of the first efforts for debugging program changes is [9]. This paper 
identifies the changes across program versions and searches among subsets of
this change set to identify which changes could be responsible for
the given observable error. In evolving program debugging, a buggy program version is simultaneously analyzed with older stable program version.

\noindent
Recently a paper proposed the DARWIN approach [10] for debugging  
program versions. DARWIN performs dynamic symbolic execution along the execution of 
a given test input in two programs. DARWIN method is basically suited for 
debugging branch errors (or code missing errors where the missing code contains 
branches). Dynamic slicing [11] has so far been studied as an aid for
program debugging and understanding. A recent work[12] uses dynamic program 
dependencies to seek the involved parts of an input that are responsible for 
a given failed output. Research such as [13],[14] combine symbolic execution and 
dependency analysis for test suite augmentation.

\noindent
[15] focuses on debugging a given failing test case based upon golden implementation driven software debugging. 
There are various published methodologies that search for failing tests (that demonstrate an observable error), 
such as -- the DSD-crasher which combines static and dynamic analysis[16], and bug finding methodology approaches relying upon software model checking
(e.g.,[17]). Symbolic execution has also been used for generating problematic
or failing tests. The work on Directed Automated Random Testing
(DART) [18] combines symbolic and concrete execution
to explore different paths of a program. A recent work [8] uses
symbolic execution on GNU Coreutilities as well as BusyBox to
compute test-suites with high path coverage. 

\noindent
Our work proposes an experimental framework for analyzing some 
published bugs in BusyBox. We employ Valgrind and Daikon in an 
attempt to locate the root causes of two memory bugs in 
BusyBox utilities. 

\section{BusyBox}
\noindent
BusyBox is a fairly comprehensive set of programs needed to run a Linux system. 
Moreover it is the de-facto standard for Embedded Linux systems, providing many standard Linux utilities, but having small code size (size 
of executables), than the GNU Core Utilities. BusyBox provides compact 
replacements for many traditional full-blown utilities found on most 
desktop and embedded Linux distributions. Examples include the file utilities 
such as ls, cat, cp, dir, head, and tail. BusyBox also provides support 
for more complex operations, such as ifconfig, netstat, route, and other 
network utilities.

\noindent
BusyBox is remarkably easy to configure, compile, and use, and it has the potential to significantly reduce the overall system resources
required to support a wide collection of common Linux utilities.
 BusyBox in general case can be built on any architecture supported by gcc.

\noindent
BusyBox is modular and highly 
configurable, and can be tailored to suit particular requirements. The
package includes a configuration utility similar to that used to configure the 
Linux kernel. The commands in BusyBox are generally simpler implementations than 
their full-blown counterparts. In 
some cases, only a subset of the usual command line options is supported. In 
practice, however, the BusyBox subset of command 
functionality is more than sufficient for most general embedded requirements.

\noindent
The BusyBox bundle functions as a single executable where the different
utilities are actually passed on at the command line for separate
invocation. It is not possible to build the individual utilities separately
and run them stand alone. For example, for running the arp utility, we need to
invoke BusyBox as {\tt busybox arp -Ainet} and record the execution trace.
Since we work on the binary level, the buggy implementation for us is
the BusyBox binary, which has a large code
base ({\em about 121000 lines of code}).

\subsection{Bugs in BusyBox}
\noindent
KLEE [8] has reported some bugs in BusyBox 1.10.2. by a test 
generation method. This paper checked all 279 BusyBox tools in series to 
demonstrate its applicability to bug finding. It has been seen that 21 bugs 
are present in BusyBox. We tried them on BusyBox version 1.4.2 and found 6 
of them still persist namely, {\em arp -Ainet, tr [, top d, printf \%Lu,
ls -co, install -m}. 

\noindent
Our objective in this paper is to locate the root causes of some of these 
bugs using memory checking and invariant analysis. In particular, we work 
with two of the 6 bugs, namely, {\em arp} and {\em top}, both of which 
result in {\em segmentation violation} on execution in BusyBox 1.4.2. To 
do so, we created a debug build of BusyBox using appropriate compiler options 
so that debugging symbols are present in the binary.
We explain our findings in the next two sections.

\section{Experiments with Valgrind}\label{sec3}
\noindent
The Valgrind tool suite provides a number of debugging and profiling tools that helps programmers 
identify memory errors in programs. Valgrind is an instrumentation framework for building dynamic 
analysis tools. It can detect many memory-related errors
that are common in C and C++ programs and that can lead to crashes and unpredictable behavior. 
Valgrind is an instrumentation framework for building dynamic analysis tools. Valgrind architecture 
is modular, so new tools can be created easily and without disturbing the existing structure. 

\noindent
Valgrind comes with a set of tools each of
which performs some kind of debugging, profiling, or similar task that helps to 
improve programs. Some of them are as below. 

\begin{itemize}

\item The default
tool that comes along with Valgrind, is {\em Memcheck}, a memory error detector.
\item {\em Cachegrind} -- a cache and branch-prediction profiler.
\item {\em Massif} --  a heap profiler.
\item {\em Helgrind} -- a thread error detector. 
\item {\em Memory leak detector}.
\item {\em Conditional jump or  uninitialized value dependent move detector}.
\item {\em Invalid Read / Write Detector}.
\end{itemize}

\noindent
Valgrind is designed to be as non-intrusive as possible. It works directly with existing executables.  
We invoke Valgrind as {\em valgrind --options executable
 program} in command line. The most important option is --tool which dictates which Valgrind tool to run. 
Regardless of which tool is in use, Valgrind takes control of the program before it starts. Debugging information is
read from the executable and associated libraries, so that error messages and other outputs can be phrased in terms of
source code locations, when appropriate. Program is then run on a synthetic CPU provided by the Valgrind core. As new code is executed,
the core hands the code to the selected tool. The tool adds its own instrumentation code to this and hands the
result back to the core, which coordinates the continued execution of this instrumented code. 

\subsection{Locating the arp bug in BusyBox}
\noindent
The {\tt arp} utility manages the kernel's network neighbor cache. It
can add or delete entries to the cache, or display the cache's current
content.  There is a bug in the BusyBox {\tt arp} implementation:
running {\tt arp} with the command-line option {\bf -Ainet} results in
a {\em segmentation fault}. 

\noindent
To isolate the root cause of this error, we 
ran the arp utility of BusyBox through Valgrind using the 
argument {-Ainet}. This results in segmentation fault when run in
BusyBox version 1.4.2, on valgrind. Figure~\ref{figbbarptest} 
shows the test results.

\begin{figure}
{\scriptsize
\centering
\begin{verbatim}
==1571== Use of uninitialized value of size 8
==1571==    at 0x4A06794: strcmp (mc_replace_strmem.c:341)
==1571==    by 0x44D7BD: get_hwtype (interface.c:936)
==1571==    by 0x444B77: arp_main (arp.c:463)
==1571==    by 0x407C08: run_applet_by_name (applets.c:489)
==1571==    by 0x407DDC: busybox_main (busybox.c:143)
==1571==    by 0x407AB7: run_applet_by_name (applets.c:480)
==1571==    by 0x407E56: main (busybox.c:72)
==1571==
==1571== Invalid read of size 1
==1571==    at 0x4A06794: strcmp (mc_replace_strmem.c:341)
==1571==    by 0x44D7BD: get_hwtype (interface.c:936)
==1571==    by 0x444B77: arp_main (arp.c:463)
==1571==    by 0x407C08: run_applet_by_name (applets.c:489)
==1571==    by 0x407DDC: busybox_main (busybox.c:143)

\end{verbatim}
}
\caption{BusyBox Arp run with valgrind} \label{figbbarptest}
\end{figure}

\begin{figure}
{\scriptsize
\begin{verbatim}
930 const struct hwtype *get_hwtype (const char *name) 
931 { 
932    const struct hwtype * const *hwp;
933    hwp = hwtypes;
934    while (*hwp != NULL) 
935    {
936      if (!strcmp((*hwp)->name, name))
937            return (*hwp);
938        hwp++;
939    }
940   return NULL;
941 }

444 int arp_main(int argc, char **argv)
445 {
446     char *hw_type;
447     char *protocol;
448 
449     /* Initialize variables... */
450     ap = get_aftype(DFLT_AF);
451     if (!ap)
452         bb_error_msg_and_die("%s: %s not 
            supported", DFLT_AF, "address family");
453 
454     getopt32(argc, argv, "A:p:H:t:i:adnDsv", 
        &protocol, &protocol,
455                  &hw_type, &hw_type, &device);
456     argv += optind;
457     if (option_mask32 & ARP_OPT_A || option_
            mask32 & ARP_OPT_p) {
458         ap = get_aftype(protocol);
459         if (ap == NULL)
460             bb_error_msg_and_die("%s: unknown 
                 %s",protocol, "address family");
461     }
462     if (option_mask32 & ARP_OPT_A || option_
            mask32 & ARP_OPT_p) {
463         hw = get_hwtype(hw_type);
464         if (hw == NULL)
465             bb_error_msg_and_die("%s: unknown 
                                       %s", 
                hw_type, "hardware type");
466         hw_set = 1;
467     }
468     //if (option_mask32 & ARP_OPT_i)... -i
469 
470     if (ap->af != AF_INET) {
471         bb_error_msg_and_die("%s: kernel only 
                      supports 'inet'", ap->name );
472     }
473 
474     /* If no hw type specified get default */
475     if (!hw) {
476         hw = get_hwtype(DFLT_HW);
477         if (!hw)
478             bb_error_msg_and_die("%s: %s not 
                supported",DFLT_HW, "hardware type");
479     }

       ....
    }
\end{verbatim}
}
\caption{BusyBox arp application} \label{fig:bbarp}
\end{figure}

\noindent
Figure~\ref{fig:bbarp} shows a fragment of the source code of {\tt
 arp} in BusyBox.  With the command-line argument {\tt -Ainet}, line
454 sets the {\tt ARP\_OPT\_A} mask in the variable {\tt
  option\_mask32}. Because no {\tt H} or {\tt t} option was given in
the command line, {\tt hw\_type} was set to NULL. The bug is at line
462: instead of checking the mask of hardware type, the program checks
for the mask of address family, {\tt ARP\_OPT\_A}, which led to line
463, which passed the NULL {\tt hw\_type} into {\tt get\_hwtype}
function, and caused a segmentation fault at line 936 due to a NULL
argument in the string comparison function {\tt strcmp}.

\subsection{Locating the top bug in BusyBox}
\noindent
The top program provides a dynamic real-time view of a running system.  
It can display system summary information as well as a list of tasks 
currently being managed by the Linux kernel. The types of system summary 
information shown and the types, order and size of information displayed 
for tasks are all user configurable and that configuration can  be  made  
persistent across restarts. 

\noindent
Top which is run with option d is buggy in busybox v1.4.2. 
If we run BusyBox top with option d, it will crash and a 
segmentation fault is reported. Option d in top must be followed 
by argument which specifies the length of the refresh delay in the 
process statistic (in seconds). The top utility should behave similarly 
when invoked with {\em -d} or {\em d}. The - is complementary in this command. 
In the normal execution of top on GNU, both {\em top d} and {\em top -d} 
wait for the parameter for refresh delay. However, BusyBox top 
behaves differently.
Figure~\ref{figbbtoptest} shows the top results.

\begin{figure}
{\scriptsize
\centering
\begin{verbatim}
==1575== Invalid read of size 1
==1575==    at 0x43310E: xstrtou_range_sfx (
                         xatonum_template.c:27)
==1575==    by 0x463605: top_main (top.c:431)
==1575==    by 0x407C08: run_applet_by_name 
				(applets.c:489)
==1575==    by 0x407DDC: busybox_main 
				(busybox.c:143)
==1575==    by 0x407AB7: run_applet_by_name 
				(applets.c:480)
==1575==    by 0x407E56: main (busybox.c:72)
==1575==  Address 0x0 is not stack'd, malloc'd 
			or (recently) free'd
==1575==    Process terminating with default action 
         		of signal 11 (SIGSEGV)
==1575==    Access not within mapped region at 
				address 0x0
==1575==    at 0x43310E: xstrtou_range_sfx 
			(xatonum_template.c:27)
==1575==    by 0x463605: top_main (top.c:431)
==1575==    by 0x407C08: run_applet_by_name 
				(applets.c:489)
==1575==    by 0x407DDC: busybox_main 
				(busybox.c:143)
==1575==    by 0x407AB7: run_applet_by_name 
				(applets.c:480)
==1575==    by 0x407E56: main (busybox.c:72)
\end{verbatim}
}
\caption{BusyBox top run with valgrind} \label{figbbtoptest}
\end{figure}

\begin{figure}
{\scriptsize
\begin{verbatim}
    414 int top_main(int argc, char **argv)
    415 {
    416     int count, lines, col;
    417     unsigned interval = 5; 
            /* default update rate is 5 seconds */
    418     unsigned iterations = UINT_MAX; 
            /* 2^32 iterations by default :) */
    419     char *sinterval, *siterations;
    420 #if ENABLE_FEATURE_USE_TERMIOS
    421     struct termios new_settings;
    422     struct timeval tv;
    423     fd_set readfds;
    424     unsigned char c;
    425 #endif /* FEATURE_USE_TERMIOS */
    426 
    427     /* do normal option parsing */
    428     interval = 5;
    429     opt_complementary = "-";
    430     getopt32(argc, argv, "d:n:b", 
	             &sinterval, &siterations);
    431     if (option_mask32 & 0x1)
                     interval = xatou(sinterval);// -d
    432     if (option_mask32 & 0x2) 
	            iterations = xato(siterations);	// -n
    433     //if (option_mask32 & 0x4) // -b
    434 
    435     /* change to /proc */
    436     xchdir("/proc");
    437 #if ENABLE_FEATURE_USE_TERMIOS
    438     tcgetattr(0, (void *) &initial_settings);
    439     memcpy(&new_settings, &initial_settings, 
                       sizeof(struct termios));
    440     /* unbuffered input, turn off echo */
    441     new_settings.c_lflag &= 
                    ~(ISIG | ICANON | ECHO | ECHONL);
    442 
    443     signal(SIGTERM, sig_catcher);
    444     signal(SIGINT, sig_catcher);
    445     tcsetattr(0, TCSANOW, 
                             (void *) &new_settings);
    446     atexit(reset_term);
    447 #endif /* FEATURE_USE_TERMIOS */
    448 
    449 #if ENABLE_FEATURE_TOP_CPU_USAGE_PERCENTAGE
    450     sort_function[0] = pcpu_sort;
    451     sort_function[1] = mem_sort;
    452     sort_function[2] = time_sort;
    453 #else
    454     sort_function = mem_sort;
    455 #endif /* FEATURE_TOP_CPU_USAGE_PERCENTAGE */
    456 
    457     while (1) {
    458         procps_status_t *p = NULL;
\end{verbatim}
}

\caption{BusyBox top application} \label{fig:bbtop}
\end{figure}

\noindent
Figure~\ref{fig:bbtop} shows a fragment of the source code of {\tt
 top} in BusyBox.  With the command-line argument {\tt d}, there is 
a crash in line 431 as reported by Valgrind. 
BusyBox's top uses a native getopt32 and checks for '-' which 
should have been ignored by top program. Going inside the getopt32 
function in Line 430, we find the code snippet shown in Figure~\ref{figg}.

\begin{figure}
\centering
{\scriptsize
\begin{verbatim}
....
if (spec_flgs & ALL_ARGV_IS_OPTS) {
   /* process argv is option, for example "ps" applet */
   if (pargv == NULL)
      pargv = argv + optind;
   while (*pargv) {
      // printf ("argv non option: %s\n", *pargv);
      c = **pargv;
      if (c == '\0') {
          pargv++;
      } 
      else {
          (*pargv)++;
          goto loop_arg_is_opt;
      }
   }
}
....
\end{verbatim}
}
\caption{BusyBox getopt32} \label{figg}
\end{figure}

\noindent
The pargv here is not NULL, thus there are some non-options to be processed. In our case here, the non-options to be processed is string 'd' and the following while-loop treat it as an option. However as soon as 'd' is treated as an option, it is no longer checked whether 'd' requires an argument. Thus the argument stored in sinterval 
in line 431 of top\_main is NULL leading to the crash.

\section{Experiments with Daikon}\label{sec4}
\noindent
Daikon is a dynamic invariant detector that reports likely program invariants.
An invariant can be defined as a property that holds at a certain point or points 
in a program. Invariants have a lot of applications. 
Some example invariants are as follows: x $>$ y,  y = 2*x-1, array $arr$ is sorted.

\noindent
The underlying approach of dynamic invariant detection is to run a program 
and observe the values computed by the program. From the observed 
values, a dynamic invariant detector tries to infer properties that were true 
over the observed executions. Daikon interacts with the 
program, firstly by obtaining the data trace by running the program under the control of a 
front end (also known as an instrumenter or tracer) that records information about variable values, 
and finally, by running the invariant detector over the data trace generated. This detects invariants 
in the recorded information generated. Daikon creates a {\em.inv}  file that contains the invariants 
in binary form. 

\subsection{Experimental Findings on arp and top using Daikon}
\noindent
We tried to run Daikon on two versions of BusyBox namely, 1.4.2 and 1.16.0. 
Our objective is to collect the invariants from the two versions when 
run on the same utilities. However, we were unsuccessful in our attempt 
due to limitations of Daikon in processing large trace files.

\noindent
Our methodology has two main steps:
\begin{enumerate} 
\item Generate the program trace using using the {\em kvasir} utility~\cite{Daikon}
\item Analyze the trace file produced above using daikon to generate the invariants.
\end{enumerate} 

\noindent
The collected traces are significantly large for both the versions. Table~\ref{tab1} 
shows a summary of our findings. During 
the trace collection phase, kvasir invokes valgrind internally and shows similar segmentation fault
as obtained by us by running Valgrind standalone. Having generated the trace files, 
we proceed to Step 2 to generate the invariants. 
Unfortunately, Daikon is unable to process the large trace files thus generated and 
aborts, without reporting any invariants. 
 
\begin{table}[!htb]
\begin{center}
{
\begin{tabular}{|c|c|c|}
\hline
Utility & Trace size (MB) & Time for Trace collection (s) \\
\hline
\hline
arp & 298, 569 & 99, 194 \\
\hline
top & 298, 570 & 92, 184 \\
\hline
\end{tabular}}
\end{center}
\caption{\em Experimental Results on BusyBox bugs using Daikon} \label{tab1}
\end{table}

\noindent
The first column of
Table \ref{tab1} is marked "Utility" --- this represents the
utility whose observable error is being diagnosed.
Each entry in the second and third columns is a tuple, where
the first entity is from BusyBox 1.4.2 and the second from
BusyBox 1.16.0.  {\em Trace Size} is the size of the trace
in Megabytes. The time for trace collection is the time required 
by kvasir to record the execution run and generate the trace. 
However, in all the cases, Daikon ends up with an error 
on the trace files thus generated and 
aborts, without reporting any invariants.
Hence, we are unable to generate the invariants. However, we tested 
our methodology on smaller examples and found diagnostic results. 
Currently, we are investigating the cause of the scalability issue 
of Daikon.

\section{Conclusion}
\noindent
In this
paper we propose a methodology for debugging errors in BusyBox, a de-facto
standard for Linux in embedded systems. Our methodology works on top of
Valgrind, a popular memory error detector and Daikon, an invariant analyzer.
We have experimented with two published errors in BusyBox with Valgrind and found promising
results. Currently, we are investigating the possibility of correlating the published bugs 
with the invariant differences after generating the invariants produced by Daikon.

\end{document}